\documentstyle [12pt,epsf] {article}

\catcode`@=12
\newcommand{\be}{\begin{equation}}
\newcommand{\ee}{\end{equation}} 
\newcommand{\ber}{\begin{eqnarray}}
\newcommand{\eer}{\end{eqnarray}} 
\newcommand{\bers}{\begin{eqnarray*}}
\newcommand{\eers}{\end{eqnarray*}}
\newcommand{\bt}{\begin{itemize}} 
\newcommand{\et}{\end{itemize}} 
\begin{document} 
\vspace{0.5in} 
\oddsidemargin -.375in 
\newcount\sectionnumber 
\sectionnumber=0 
\def\bra#1{\left\langle #1\right|} 
\def\ket#1{\left| #1\right\rangle} 
\def\be{\begin{equation}} 
\def\ee{\end{equation}} 
\thispagestyle{empty} 
\begin{flushright} 
UdeM-GPP-TH-02-98 \\
UTPT-02-08\\
WIS/21/02-JUNE-DPP\\
TAUP 2706-02\ 
\end{flushright} 
\vspace {.5in} 
\begin{center} {\Large\bf Non Standard $\eta-\eta^{\prime}$ mixing and the 
Nonleptonic $B$ and $\Lambda_b$ 
Decays to $\eta$ and $\eta^{\prime}$ \\}
\vspace{.5in}
{{\bf Alakabha Datta{\footnote{email: datta@lps.umontreal.ca}}${}^{a}$}, 
{\bf Harry J. Lipkin {\footnote{email: 
harry.lipkin@weizmann.ac.il}}${}^{b}$} and
{\bf Patrick. J. O'Donnell {\footnote{email: pat@medb.physics.utoronto.ca}}${}^{c}$} 
 \\}
\vspace{.1in}
${}^{a}$
{\it Laboratoire Ren\'e J.-A. L\'evesque, Universit\'e de Montr\'eal,}
\\ {\it C.P. 6128, succ.\ centre-ville, Montr\'eal, QC, Canada H3C
  3J7} \\ ${}^{b)}$ {\it Department of Particle Physics,\\ Weizmann
  Institute,\\ Rehovot 76100, Israel \\and\\ School of Physics and
  Astronomy, \\ Tel-Aviv University,\\ Tel-Aviv 69978, Israel \\ 
  ${}^{c}$ {\it Department of Physics and Astronomy,\\ University of
    Toronto, Toronto, Canada.}\\ } \end{center}

\begin{abstract}
  Radial mixing in the pseudoscalar $\eta-\eta^{\prime}$ systems can
  be generated from hyperfine interactions and annihilation terms.
  For the $\eta-\eta^{\prime}$ system we find the effects of radial
  mixing are appreciable and seriously affect the decay branching
  ratios for $B \rightarrow \eta(\eta^{\prime})K(K^*)$, mainly by
  modifying the $ B \rightarrow \eta(\eta^{\prime})$ form factors.  In
  particular, the effect of radial mixing in conjunction with the
  interference effects among penguin amplitudes can resolve puzzles in
  the $B \rightarrow \eta(\eta^{\prime}) K$ decays.  The decay
  $\Lambda_b \rightarrow \Lambda \eta(\eta^{\prime})$ on the other
  hand is dominated by a single amplitude so that the significant
  interference effects of $B$ decays are absent here. Moreover, since
  no $\Lambda_b \rightarrow \eta(\eta^{\prime})$ form factors are
  involved here, the effect of radial mixing is essentially
  negligible. Hence, unlike the $B$ system, we do not predict a large
  enhancement of $\Lambda_b \rightarrow \Lambda \eta^{\prime}$
  relative to $\Lambda_b \rightarrow \Lambda \eta$.

\end{abstract}



\begin{figure}[htb] 
   \centerline{\epsfysize 4.2 truein \epsfbox{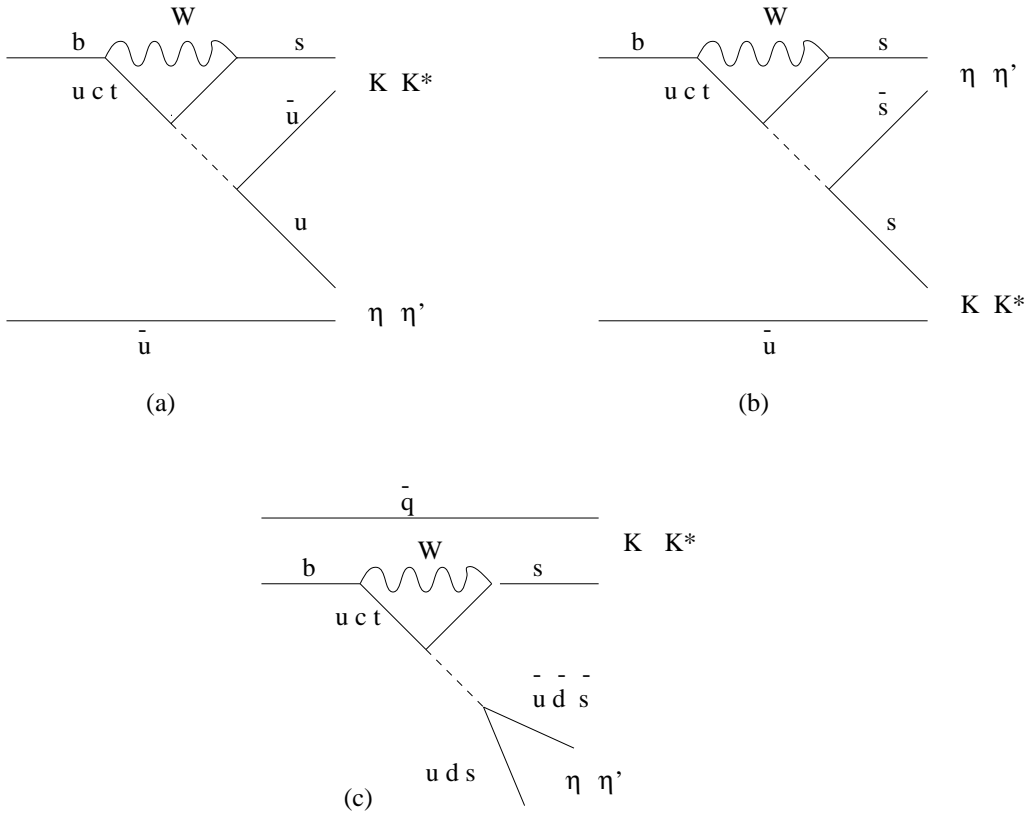}} 
   \caption{The  diagrams contributing to $B \to \eta(\eta^{\prime})
     K(K^*)$ decays. The dashed line represents a gluon or a $\gamma$
     or a Z boson. Tree diagrams are not shown. For the decays
     $\Lambda_b \rightarrow \Lambda \eta(\eta^{\prime})$ diagram (a)
     is absent.}
\label{eta} 
\end{figure}

Nonleptonic $B$ decays play a very important role in the study of CP
violation. It is expected that these will test the standard model (SM)
picture of CP violation or provide hints for new physics.  Some clues
to possible new physics may be given by recent experimental data for
$B$ and $D$ decays into final states containing the $\eta$ and
$\eta^{\prime}$ pseudoscalar mesons.  So far these have remained
unexplained in the standard treatments.  One possible source of this
disagreement comes from the fact that most models describe the $\eta$
and $\eta^{\prime}$ mesons as node-less ground-state s-wave $q \bar q$
systems. In a previous paper \cite{Vector}, we have shown that in a
factorization approximation, suitable for the treatment of two body
decays, radial excitations are favored. We will follow this treatment
for the pseudoscalar mesons and consider the $\eta$ and
$\eta^{\prime}$ wave functions to be mixtures of the ground state and
radially excited $q \bar q$ states.  This will alter the high momentum
behavior of the $\eta$ and $\eta^{\prime}$ wave functions.  So, for
example, for $B \to \eta(\eta^{\prime})K(K^*)$ decays, as shown in
Fig.~\ref{eta} (a), factorization results in the kaon leaving the weak
vertex with its full momentum; the remaining quark carries the full
momentum of the final $\eta(\eta^{\prime})$ meson.  A large internal
momentum transfer is needed to hadronize this quark with the spectator
anti-quark to form an $\eta$ or $\eta^{\prime}$ final state.  This
will favor radial excitations since they have a much higher mean
internal momentum.  On the other hand, there is no such diagram in the
decay $\Lambda_b \rightarrow \Lambda \eta(\eta^{\prime})$ and we
should not expect any similar radial enhancement of the form factors.

The most general description of $\eta-\eta^{\prime}$ system involves
four different radial wave functions and cannot be described by
diagonalizing a simple $2 \times 2$ matrix with a single mixing angle
\cite{Mixing}. The normalized $\eta-\eta^{\prime}$ wave functions may
be written in terms of two angles as

\ber
\ket{\eta} & = & \cos{\phi}\ket{N} -\sin{\phi}\ket{S},\nonumber\\
\ket{\eta^{\prime}} & = & \sin{\phi^{\prime}}\ket{N^{\prime}} +
\cos{\phi^{\prime}}\ket{S^{\prime}}, 
\label{generalmost}
\eer
where $\ket{N}$, $\ket{N^{\prime}}$, $\ket{S}$ and $\ket{S^{\prime}}$
are arbitrary isoscalar non--strange and strange quark-antiquark wave
functions, respectively. In the traditional picture where the
$\eta-\eta^{\prime}$ mixing is described by a single mixing angle,
\ber 
\ket{N} & = & \ket{N^{\prime}},\nonumber\\ 
\ket{S} & = & \ket{S^{\prime}},\nonumber\\ 
\phi & = & \phi^{\prime}. \label{std} 
\eer
Some theoretical input \cite{Feldmann} is needed to obtain the mixing
angle $\phi$ from experimental data; tests of the standard mixing case
involving $B$ decays have been discussed in Ref. \cite{Mixing}.

In the absence of a model independent extraction of $\phi$ we will use
as our standard mixing the Isgur mixing \cite{Isgur}
\ber
\ket{\eta}_{std} & = & \frac{1}{\sqrt{2}}\left[N_0 -S_0 \right]
\nonumber\\ 
\ket{\eta^{\prime}}_{std} & = & \frac{1}{\sqrt{2}}\left[N_0 +S_0
\right] \ \label{Isgur} 
\eer
which corresponds to the mixing angle $\phi=45^0$. We will treat the
mixing given here as ``standard'' in the sense that we will compare
results calculated with non--standard mixing coming from radially
excited components in the $\eta(\eta^{\prime})$ wave functions with
those given in Eq. (\ref{Isgur}).

To obtain the eigenstates and eigenvalues in the $\eta-\eta^{\prime}$
system, including radial excitations, we diagonalize the mass matrix
\ber
< q_a'\bar{q}_b',n'|M|q_a \bar{q}_b,n> & = & 
\delta_{aa'} \delta_{bb'} \delta_{nn'}
(m_a +m_b +E_n) +\delta_{aa'}\delta_{bb'}\frac{B}{m_am_b} 
{\vec{s}_a \cdot \vec{s}_b} \psi_n(0)\psi_{n'}(0)
\nonumber\\
& + & \delta_{ab} \delta_{a'b'} \frac{A}{m_am_b} 
\psi_n(0)\psi_{n^{\prime}}(0).
\label{massmatrix}
\eer
where $\vec{s}_{a,b}$ and $m_{a,b}$ are the quark spin operators and
masses.  Here $n=0,1,2$ and the basis states for the isoscalar mesons
are chosen as 
 $|N>_n=| u \bar{u} + d \bar{d}>_n/\sqrt{2}$ and
$|S>_n=|s \bar{s}>_n$ for the  non--strange and strange wavefunctions. 
The excitation energy of the $n^{th}$
radially excited state is $E_n$.

This has the same structure as that of the vector system \cite{Vector}
involving the $\omega - \phi$ mesons. The only differences between the
pseudoscalar and vector mass matrices are in the values of the
parameters, which are determined by fitting the experimental masses.
 
The mixing of the strange and non--strange components in the $\eta$
and $\eta^{\prime}$ wave functions is assumed usually to come from a
short-range flavor-singlet interaction which is absent for the pion
and kaon. This was shown to have a negligible effect on the vector
mesons \cite{Vector}.  This flavor mixing contribution was called the
``annihilation" contribution \cite{georgi} in earlier papers on
$\eta-\eta^{\prime}$ mixing.  This phenomenological contribution, $A$,
which we shall continue to refer to as an annihilation contribution,
has no direct relation to the annihilation diagrams of QCD.  The QCD
origin of flavor mixing in the $\eta-\eta^{\prime}$ system is not
fully understood and probably has to do with gluon effects and the
non--trivial structure of the QCD vacuum.  We will assume that these
effects are included in an effective parameter $A$. When the parameter
A=0, we see that in the vector case the $\rho$ and $\omega$ masses are
equal while in the pseudoscalar case the $\pi$ and $\eta$ masses are
equal. Thus, fitting the experimentally small $\rho - \omega$ mass
difference and the experimentally large $\pi - \eta$ mass difference
immediately requires a small value for A and small flavor-mixing for
the vectors and a large value for A and large flavor-mixing for the
pseudoscalars.

In addition to this flavor--mixing interaction we also include a
short--range hyperfine interaction with a strength $B$. As in
\cite{Vector} we have allowed the annihilation term to have a flavor
dependence from mass factors and the wave functions, modeling it on
the hyperfine interaction \cite{Cohen:1979ge}. Evidence for flavor
dependence of the annihilation term in $\eta-\eta^{\prime}$ mixing has
been known for some time \cite{jackson}.  Even without knowledge of
the QCD origin of this term we can see that it may be considered as a
short-range repulsion which acts only on a singlet state, with the
matrix element vanishing for the octet state in the $SU(3)$ limit.  In
a simple quark potential model this implies that the singlet and octet
states have different radial wave functions.

These effects break nonet symmetry and necessarily correct all results
which assume that members of the pseudoscalar nonet all have the same
radial wave function.  It is this non--standard $\eta(\eta^{\prime})$
mixing that has important implications for the nonleptonic decays $B
\to \eta(\eta^{\prime})K(K^*)$.

In a quark potential model, flavor-SU(3) breaking by quark masses
makes the kinetic energies of the strange and non--strange components
of the pseudoscalars different. Thus the $\eta-\eta^{\prime}$
system has now four different radial wave functions and is no longer
described by a simple $2 \times 2$ matrix with one mixing angle.  Our
approach here is to follow the example of Refs.
\cite{Cohen:1979ge,Frank:1984bj} and choose a basis for our radial
wave functions. In this basis the difference in radial wave functions
of the strange and non--strange parts of the $\eta-\eta^{\prime}$
system is described by admixtures of radial excitations. This allows
us to write the normalized $\eta-\eta^{\prime}$ wave functions as
\ber 
\ket{\eta} & = & a_1 \ket{N} +a_2 \ket{S},\nonumber\\ 
\ket{\eta^{\prime}} & = & a_1^{\prime} \ket{N^{\prime}} 
+a_2^{\prime} \ket{S^{\prime}},\
\label{general}
\eer
where the normalized states $\ket{N}$, $\ket{N^{\prime}}$, $\ket{S}$
and $\ket{S^{\prime}}$ have the general structure 
\ber 
\ket{N} & = & \frac{\ket{N}_0 +b_1\ket{N_1}+b_2\ket{N_2}} {\sqrt{1+b_1^2+b_2^2}},  \nonumber\\  
\ket{N^{\prime}} & = & \frac{\ket{N}_0 +b_1^{\prime}\ket{N_1}+
  b_2^{\prime}\ket{N_2}} {\sqrt{1+b_1^{\prime 2}+b_2^{2 \prime}}},
\nonumber\\ 
\ket{S} & = & \frac{\ket{S}_0 +c_1\ket{S_1}+c_2\ket{S_2}} {\sqrt{1+c_1^2+c_2^2}}, 
\nonumber\\  
\ket{S^{\prime}} & = & \frac{\ket{S}_0 +c_1^{\prime}\ket{S_1}+
  c_2^{\prime}\ket{S_2}} {\sqrt{1+c_1^{\prime 2}+c_2^{2 \prime}}},\
\label{ketgeneral} 
\eer  
with $|N>_{0,1,2}=| u \bar{u} + d \bar{d}>_{0,1,2}/\sqrt{2}$ and
$|S>_{0,1,2}=|s \bar{s}>_{0,1,2}$ representing the various non--strange
and strange radial excitations. The normalization of the
$\eta-\eta^{\prime}$ wave functions in Eq.~(\ref{general}) then leads
to $a_1^2+a_2^2 = a_1^{\prime 2} +a_2^{\prime 2} = 1$.

The values of $A$ and $B$ in the mass matrix, Eq.~(\ref{massmatrix}),
are fitted from the measured masses.  We use the phase convention of
Ref \cite{Frank:1984bj} where the wave functions at the origin in
configuration space, which enter in the hyperfine and annihilation
terms in the mass matrix, are positive (negative) for the even (odd)
radial excitations. The mass matrix is a $6 \times 6$ matrix which is
diagonalized to give the six masses and mixings. However, for our
purposes we will only need the predictions for the $\eta$ and
$\eta^{\prime}$ masses and wave functions. Several solutions that give
acceptable values of the masses can be obtained. We choose those
solutions for the linear, quadratic and quartic confining potentials
that are similar in predictions for the $\eta(\eta^{\prime})$ masses.

With $B=0.065m_u^2$ we obtain the eigenvalues and
eigenstates in Table~\ref{etalinear} for the linear potential, in
Table~\ref{etaquadratic} for the harmonic potential and in
Table~\ref{etaquartic} for the quartic potential. Our results for the
harmonic potential are similar to those of Ref \cite{Frank:1984bj}
where a slightly different mass mixing matrix has been used to obtain
the $\eta-\eta^{\prime}$ mixing.

\begin{table}[thb]
\caption{Eigenvalues and eigenstates for the $\eta-\eta^{\prime}$
  system with $A=0.045m_u^2$ } \begin{center} \begin{tabular}{|c|c|c|c|c|c|c|c|}
\hline
Linear  & $N_0$ & $N_1$ & $N_2$ & $ S_0$ & $S_1$& $ S_2$\\ \hline $\eta(0.544)$ & 0.961 &-0.198& 0.108 & -0.150 & 0.050& -0.032 \\ \hline 
$\eta^{\prime}(0.924)$ &
0.170 &0.039& -0.016 & 0.974 &-0.126 &0.049 \\
\hline 
\end{tabular}
\end{center}
\label{etalinear}
\end{table}

\begin{table}[thb]
\caption{Eigenvalues and eigenstates for the $\eta-\eta^{\prime}$ system with $A=0.065m_u^2$ } \begin{center} \begin{tabular}{|c|c|c|c|c|c|c|c|}
\hline
Harmonic  & $N_0$ & $N_1$ & $N_2$ & $ S_0$ & $S_1$ & $ S_2$\\ \hline $\eta(0.547)$ & 0.913 &-0.252& 0.154 & -0.249 & 0.107& -0.076 \\ \hline 
$\eta^{\prime}(0.931)$ &
0.316 &0.109& -0.049 & 0.925 &-0.148 &0.088 \\
\hline 
\end{tabular}
\end{center}
\label{etaquadratic}
\end{table}

\begin{table}[thb]
\caption{Eigenvalues and eigenstates for the $\eta-\eta^{\prime}$ system with $A=0.11m_u^2$ } \begin{center} \begin{tabular}{|c|c|c|c|c|c|c|c|}
\hline
Quartic  & $N_0$ & $N_1$ & $N_2$ & $ S_0$ & $S_1$ & $ S_2$\\ \hline $\eta(0.547)$ & 0.764 &-0.287& 0.198 & -0.441 & 0.248& -0.196 \\ \hline 
$\eta^{\prime}(0.940)$ &
0.623 &0.350& -0.177 & 0.658 &-0.134 &0.087 \\
\hline 
\end{tabular}
\end{center}
\label{etaquartic}
\end{table}

The entries for $\eta(\eta^{\prime})$ mixing in
Tables (\ref{etalinear}, \ref{etaquadratic} and \ref{etaquartic}) are
the coefficients for the eigenstates in Eqs. (\ref{general},
\ref{ketgeneral}).  These are sensitive to the confining potential;
there can be substantial radial mixing, affecting the predictions for
$B \to \eta(\eta') K(K^*)$ decays. In going from linear to quadratic
to quartic potential the $\eta(\eta')$ mixing deviates more from the
ideal mixing case. This corresponds to the increase in value of $A$.

Let $\ket {\eta}_g$ and $\ket {\eta'}_g$ represent the (unnormalized)
portions of the physical wave functions that are in the ground state
configuration in Tables (4-6).  These states may then be written in
terms of the ``standard'' $\eta-\eta^{\prime}$ states defined in Eq.
(\ref{Isgur}). For the linear potential we find,
\ber
\ket{\eta^{\prime}}_g & = 
& 0.81\ket{\eta^{\prime}}_{std} -0.57 \ket{\eta}_{std}, \nonumber\\ 
\ket{\eta}_g & = & 0.79\ket{\eta}_{std} +0.57
\ket{\eta^{\prime}}_{std}.\ 
\eer 
For the harmonic potential we find, 
\ber \ket{\eta^{\prime}}_g & = & 0.88\ket{\eta^{\prime}}_{std} -0.43
\ket{\eta}_{std}, \nonumber\\ 
\ket{\eta}_g & = & 0.82\ket{\eta}_{std} +0.47
\ket{\eta^{\prime}}_{std}.
\ \eer  
For the quartic potential we find, 
\ber \ket{\eta^{\prime}}_g & = & 0.91\ket{\eta^{\prime}}_{std} -0.025
\ket{\eta}_{std}, \nonumber\\ 
\ket{\eta}_g & = & 0.85\ket{\eta}_{std} +0.23
\ket{\eta^{\prime}}_{std}.\ 
\eer 
Thus all three confining potentials give mixings for the
$\eta-\eta^{\prime}$ that have substantial overlap with the ``standard
mixing". The mixing from the quartic potential is closest to the
standard mixing in the sense that here one has the smallest component
of the $\ket{\eta}_{std} (\ket{\eta^{\prime}}_{std}$ in
$\ket{\eta^{\prime}}_g (\ket{\eta}_g)$. Several tests of non--standard
mixing have been discussed in Ref\cite{Mixing}. Of particular interest, for
non leptonic $B$ decays, are the ratios defined in \cite{Mixing} as
\ber
r_d & \equiv &\frac
{p_{\eta^{\prime}}^3\Gamma( {\bar B}^0 \to J/\psi \eta) }
{p_{\eta}^3\Gamma( {\bar B}^0 \to J/\psi \eta^{\prime})} \
\eer
and
\ber
r_s & \equiv &\frac
{p_{\eta^{\prime}}^3\Gamma( {\bar B_s}^0 \to J/\psi \eta) }
{p_{\eta}^3\Gamma( {\bar B_s}^0 \to J/\psi \eta^{\prime})}.\
\eer
We then have the prediction, for the standard mixing,
\ber
r & = & \sqrt{r_dr_s}=\sqrt{\cot^2\phi \tan^2\phi}=1.\
\label{rnl}
\eer 
On the other hand, with the non--standard mixing, the ratio $r$ 
can be quite different from unity. In particular, with the mixing in 
Table \ref{etaquartic}, we found that $r$ could be as low as
 $ \sim 0.2$ \cite{Mixing}. 

Having now defined how we will construct the $\eta(\eta^{\prime})$
wave functions we now consider $B \to \eta(\eta^{\prime})K(K^*)$
decays. These are dominated by the penguin diagrams shown in 
Fig. \ref{eta} since the tree term is color and CKM \cite{CKM}
suppressed.

In Fig.~\ref{eta} (a) factorization results in the kaon leaving the
weak vertex with its full momentum; the remaining quark combines with
the spectator quark to form the final $\eta(\eta^{\prime})$ meson.
Fig. \ref{eta} (b) shows the $\bar{s}$ quark in the QCD penguin
combining with the $s$ quark from the $ b \to s$ transition to form
the $\eta(\eta^{\prime})$. Another possibility is shown in Fig.
\ref{eta} (c) in which a $q\bar{q}$ pair (where $ q=u,d, s$) appearing
in the same current in the effective Hamiltonian, hadronizes to the
$\eta(\eta^{\prime})$.  This term is often referred to as being OZI
suppressed \cite{Okubo:1963fa,Zweig} with respect to the other terms
in the decay amplitude. However, this may not be the case 
for $\eta^{\prime}$ in the final state
because the OZI suppressed terms add constructively while, for
the $\eta$ in the final state the OZI suppressed terms tend to cancel
among themselves.  

We expect the OZI suppressed terms to be more important in $B \to K P$
than in $B \to K V$ decays where $P$ is a pseudoscalar and $V$ is a
vector state.  This follows from the fact that in $J/\psi$ and
$\Upsilon$ decays we know that the OZI-forbidden process requires
three gluons for coupling to a vector meson and two gluons for
coupling to a pseudoscalar. Thus the contribution of the OZI
suppressed term should be much smaller in the $B \to K \rho^0(\omega)$
and $B \to K \phi$ decays than in $ B \to K \eta$ and $B \to K
\eta^{\prime}$ decays \cite{Lipkin:1995hn, Lipkin:1997ke}.

If it is assumed that the OZI terms are forbidden then definite
predictions about the branching ratios $B \to \eta K/B \to
\eta^{\prime} K$ and $ B \to \eta K^*/B \to \eta^{\prime} K^*$
are possible \cite{Lipkin:1991us,Lipkin:1992fd}.  Here we will first
derive these predictions using the factorization assumption and the
standard $\eta-\eta^{\prime}$ mixing in Eq. (\ref{Isgur}).  We will
then study how the predictions change with non--standard mixing for
$\eta-\eta^{\prime}$ and with the inclusion of the OZI terms.  For $B$ decays,
we define the two ratios
\ber 
R_{K} & = & BR(B^- \rightarrow K^- \eta )/BR(B^- \rightarrow K^- \eta^{\prime})
\label{bk} 
\eer  
and 
\ber R_{K^*} & = & 
BR(B^- \rightarrow K^{-*} \eta )/BR(B^- \rightarrow K^{-*} \eta^{\prime}),
\label{bkstar} 
\eer 
and, for $\Lambda_b \rightarrow \Lambda \eta(\eta^{\prime})$, define the
ratio 
\ber R_{\Lambda} & = & BR(\Lambda_b \rightarrow \Lambda \eta )/
BR(\Lambda_b \rightarrow \Lambda \eta^{\prime}).
\label{lambda}
\eer
It has been shown \cite{Lipkin:1991us,Lipkin:1992fd} that there is a parity
selection rule for the decays $B \to \eta(\eta^{\prime})K^{(*)}$. This
fixes the relative phase between the penguin amplitudes, Fig.
\ref{eta}(a) and Fig. \ref{eta}(b), for the strange and non--strange
contributions to the $\eta$ and $\eta^{\prime}$ final states.  In
particular, the parity selection rule predicts that there is a relative
sign difference between the strange and non--strange penguin amplitudes
in $B \to K \eta(\eta^{\prime})$ and $B \to K^* \eta(\eta^{\prime})$.
 
With the use of factorization and this parity selection rule,
we find
\ber
R_{K} & \approx & \left|\frac{f_KF^{+}_{\eta}+ f_{\eta}^sF^{+}_{K}}
  {f_KF^{+}_{\eta^{\prime}}+ f_{\eta^{\prime}}^sF^{+}_{K}}\right|^2,\nonumber\\
R_{K^{*}} & \approx & \left|\frac{f_{K^{*}}F^{+}_{\eta}- f_{\eta}^sF^{+}_{K^{*}}} {f_{K^{*}}F^{+}_{\eta^{\prime}}- 
f_{\eta^{\prime}}^sF^{+}_{K^{*}}}\right|^2,\nonumber\\
R_{\Lambda} & \approx &
\left|\frac{f_{\eta}^sF_{\Lambda}} 
{f_{\eta^{\prime}}^sF_{\Lambda}}\right|^2,\
\label{rk}
\eer
where we have set the masses of the pseudoscalars in the final states
to zero and introduced the form factor $F^{+}$ \cite{BSW}. For 
$\Lambda_b$ decays the form factors $F_{\Lambda}$ as well as the heavy
quark relations between them can be found in Ref. \cite{Dattaff};
we have also neglected  terms of $O(m_{\Lambda}/m_{\Lambda_b})$.  
In the
above equation $f_K$ is the kaon decay constant and the decay
constants $f_{\eta}^q$ and $f_{\eta^{\prime}}^q$ are defined by
\ber 
if_{\eta(\eta^{\prime})}^q p^{\mu}_{\eta(\eta^{\prime})} & = & \bra{\eta(\eta^{\prime})}
\bar{q}\gamma^{\mu}(1-\gamma_5) q \ket{0}.\ 
\label{decayconstant} 
\eer 
{}For standard mixing, Eq. (\ref{Isgur}), and assuming $SU(3)$ flavor
symmetry, we obtain the following relations between the various
form factors and decay constants appearing in Eq. (\ref{rk}).
\ber
f_{\eta}^{u,d} & \approx & \frac{f_K}{2}  \quad;
f_{\eta}^s \approx \frac{-f_K}{\sqrt{2}} \quad;
f_{\eta^{\prime}}^{u,d} \approx \frac{f_K}{2}\quad;
f_{\eta^{\prime}}^s \approx \frac{f_K}{\sqrt{2}}\nonumber\\
F_{\eta}^{+} & \approx & \frac{F_K^{+}}{{2}}\quad ;
F_{\eta^{\prime}}^{+} \approx \frac{F_K^{+}}{{2}}.\ 
\label{fsymmetry} 
\eer 

{}From Eqs. (\ref{rk}, \ref{fsymmetry}) we find
\ber
R_K & \approx & \left|\frac{\frac{1}{2}-\frac{1}{\sqrt{2}}}
{\frac{1}{2}+\frac{1}{\sqrt{2}}}\right|^2 \sim 0.03. \ 
\label{rkn} 
\eer 
Hence there is substantial interference between the two penguin
amplitudes. This gives a small value for $R_K$, consistent with the
current experimental limits.  Neglecting differences in form factors
between $B \to P$ to $B \to V$ transitions, we find
\ber
R_{K^*} & \approx & \left|\frac{\frac{1}{2}+\frac{1}{\sqrt{2}}}
{\frac{1}{2}-\frac{1}{\sqrt{2}}}\right|^2 \sim \frac{1}{R_{K}}=33. \ 
\label{rkstarn} 
\eer
{}For the $\Lambda$ ratio, only Fig. \ref{eta}(b) contributes so that
\ber
R_{\Lambda} &\sim & 1\
\label{rlambdan}
\eer
We now calculate the ratios $R_{K(K^*)}$ and $R_{\Lambda}$ with a
nonstandard $\eta-\eta^{\prime}$ mixing. This is a more complicated
calculation and so we will use the mixing for a quartic potential,
given in Table~(\ref{etaquartic}), as an example, since the ground
states of this mixing are those most similar to the standard mixing,
Eq.  (\ref{Isgur}).

A key ingredient in the parity selection rule that predicts the
relative phases of the strange and non--strange contributions to $\eta$
and $\eta^{\prime}$ final states is approximate flavor symmetry. In
the flavor symmetry limit the radial wave functions of $\pi$, $K$,
$\eta$ and $\eta^ {\prime}$ (up to mixing factors in $\eta
(\eta^{\prime}$)) are all the same. Since all are ground state wave
functions with no nodes and a constant phase over the entire radial
domain, flavor symmetry breaking may change the radial shape and size
of the wave function but will not change the relative phases. However,
when radially excited wave functions, with nodes that differ from the
ground state wave functions, are also considered, there may be
complicated changes in the relative phases.

For example, with factorization, B decay into the non--strange part of
the $\eta$ and $\eta^{\prime}$ involves a point--like form factor for
the kaon and and a hadronic overlap integral for the $\eta
(\eta^{\prime})$.  On the other hand, B decay into the strange part of
the $\eta$ and $\eta^{\prime}$ involves a point--like form factor for
the $\eta (\eta^{\prime})$ and a hadronic overlap integral for the
kaon.  A phase ambiguity can therefore arise between the two penguin
amplitudes coming from the hadronic overlap integral for the $\eta$
and $\eta^{\prime}$ since it involves a mixing among different parts
of wave functions.

One way to simulate flavor symmetry breaking would be to define, for
the $\eta (\eta^{\prime})$, an effective wave function
\ber
\Psi_{eff} & = & a_0 \Psi_{N_0} + a_1 r_1 \Psi_{N_1} +a_2 r_2
\Psi_{N_2}\ 
\label{ewave function} 
\eer 
where the $a_i$'s are from Table \ref{etaquartic} and $r_1$ and $r_2$,
representing flavor breaking effects, can have either sign; form
factors can then be obtained from this effective wave function.  We
will consider two choices of $r_1$ and $r_2$. In the first, and
simplest, case we set $r_1=r_2=1$. Otherwise, we choose $r_1 = - r_2$ 
so that the contributions from the various radial
excitations and the ground state add constructively. 

The problem of calculating form factors for
 transitions to radially excited states warrants
a separate investigation of its own and is beyond the scope of this
work. Therefore for a decay to a radially excited state $M^{\prime}$ we  use
for the form factor $B \to M^{\prime}$
$$B\to M^{\prime}= (B \to M^{\prime}/B \to M) \times B \to M.$$
The first factor in the right hand side of the above, $B \to M^{\prime}/B
\to M$, is obtained from a constituent model \cite{Aleksan:1995bh}.
While this is a simplified quark model that is not expected to
correctly predict the absolute values of the form factors, it does
correctly exhibit the scaling properties of the form factors expected
from heavy quark symmetry. We use this model to predict the
ratio $B \to M^{\prime}/B \to M$ assuming that corrections to the form
factors, calculated in a more complete model, cancel in the ratio. The
second factor $B \to M$, representing the form factors for
transition from $B$ to the ground state $M$, 
is taken from Ref. \cite{BSW}. The
decay constants in Eq. (\ref{decayconstant}) are calculated
from the wave functions for the quartic potential using the formula in 
Ref \cite{DattaAlaa}

For the $F_{+}$ form factor we find
\ber 
\frac{F_{+nonstandard}^{\eta^{\prime}}}{F_{+standard}^{\eta^{\prime}}}
&\approx & 1.5,1.7 \quad;\,\, \frac{F_{+nonstandard}^{\eta}}{F_{+standard}^{\eta}}
\approx  0.5, 2.1 ,\
\label{formfactor1}
\eer
where $F_{+standard}$ and $F_{+nonstandard}$ are the form factors
calculated in the standard mixing in Eq. (\ref{Isgur}), and for the
non--standard mixing in Table (\ref{etaquartic}).  The two numbers in
the equation above correspond to the two choices for $r_{1,2}$
mentioned above. The form factor with non--standard mixing
does not change much for the $\eta^{\prime}$ but changes significantly
for the $\eta$ for the two choices of $r_{1,2}$.  {}For the decay
constants we find
\ber \frac{f^{u,d}_{{\eta}_{nonstandard}}}{f^{u,d}_{{\eta}_{standard}}} & \approx & 1\quad; \frac{f^{u,d}_{{\eta^{\prime}}_{nonstandard}}}
{f^{u,d}_{{\eta^{\prime}}_{standard}}}  \approx  1.1\quad; \frac{f^{s}_{{\eta}_{nonstandard}}}{f^{s}_{{\eta}_{standard}}}  \approx 
 0.8\quad; \frac{f^{s}_{{\eta^{\prime}}_{nonstandard}}}{f^{s}_{{\eta^{\prime}}_
{standard}}}  \approx  1.2.\
\label{dconstant1}
\eer
Using the second entry in Eq.~\ref{formfactor1} we then have, approximately,
\ber \frac{F_{+nonstandard}^{\eta^{\prime}}}{F_{+standard}^{\eta^{\prime}}}
&\approx & 2.0 \approx \frac{F_{+nonstandard}^{\eta}}{F_{+standard}^{\eta}},\nonumber\\
{f^{u,d,s}_{{\eta}_{nonstandard}}}& \approx & {f^{u,d,s}_{{\eta}_{standard}}} 
\quad;
{f^{u,d,s}_{{\eta^{\prime}}_{nonstandard}}}  \approx 
{f^{u,d,s}_{{\eta^{\prime}}_{standard}}}. \
\label{new}
\eer

Together with the relations in Eq. (\ref{fsymmetry}) 
we obtain for the
non--standard $\eta-\eta^{\prime}$ mixing in Eq. (\ref{new}) 
\ber 
f_{\eta_{nonstandard}}^{u,d} & \approx & f_{\eta_{standard}}^{u,d}
\approx \frac{f_K}{2}\quad;\,\, 
f_{\eta_{nonstandard}}^s \approx f_{\eta_{standard}}^s \approx 
\frac{-f_K}{\sqrt{2}}; \nonumber\\ 
f_{\eta^{\prime}_{nonstandard}}^{u,d} & \approx & f_{\eta^{\prime}_{standard}}^{u,d} 
\approx \frac{f_K}{2}\quad;
f_{\eta^{\prime}_{nonstandard}}^s \approx 
f_{\eta^{\prime}_{standard}}^s \approx
\frac{f_K}{\sqrt{2}};\nonumber\\ 
F_{\eta_{nonstandard}}^{+} &\approx& 2F_{\eta_{standard}}^{+}  \approx  
{F_K^{+}}; \nonumber\\
F_{\eta^{\prime}_{nonstandard}}^{+}& \approx &  2F_{\eta^{\prime}_{standard}}^{+} 
\approx {F_K^{+}}.\
\label{newfsymmetry}
\eer
{}From Eq. (\ref{newfsymmetry}) it can easily be checked that the
predictions $R_{K}$ and $R_{K^*}$ in Eqs.  (\ref{rkn} and
\ref{rkstarn}) remain unchanged under a scaling by a factor of two
(last parts of Eq. (\ref{newfsymmetry})). Since $R_{\Lambda}$ does not
depend on $B \rightarrow \eta(\eta^{\prime})$ form factors and the
decay constants change little with non--standard mixing the prediction
for $R_{\Lambda}$ also remains essentially unchanged with non--standard
$\eta(\eta^{\prime})$ mixing.

It has been shown, by one of the authors of this paper, that flavor
topology characteristics of charmless $B$ decays give rise to
additional sum rules connecting $B$ decays to $K
\eta({\eta^{\prime}})$ and $ K \pi$ final states. One of the
interesting sum rules \cite{Lipkin:1997ke,Lipkin:1997ad,
Lipkin:2000sf} is, neglecting phase space corrections,
\ber
R & = & \frac{\Gamma[B^{\pm} \to K^{\pm} \eta^{\prime}]
+
\Gamma[B^{\pm} \to K^{\pm} \eta]}
{\Gamma[B^{\pm} \to K^{\pm} \pi^0]}  \le  3 \
\label{sumrule}
\eer
Recent experimental measurements \cite{Chen:2000hv}, however, show the
above sum rule to be invalid.

With the same assumptions that led to Eq. (\ref{rk}) 
\ber 
R & \approx & \frac{\left|f_KF^{+}_{\eta}+ f_{\eta}^sF^{+}_{K}\right|^2+ \left|f_KF^{+}_{\eta^{\prime}}+ f_{\eta^{\prime}}^sF^{+}_{K}\right|^2}
{\left|F^{+}_{\pi^0}f_K\right|^2}.
\label{pred0}
\eer
With standard mixing this gives $R \approx $ 3.  However, using the
non--standard $\eta-\eta^{\prime}$ mixing of Eq. (\ref{newfsymmetry}),
leads to $R \approx $ 6, a value which is consistent with experiment.
A similar result occurs for $K^*$ final states. In particular, we have
\ber 
\frac{\Gamma[B^{\pm} \to K^{*\pm} \eta]} {\Gamma[B^{\pm} \to K^{*\pm}
  \pi^0]}  & \approx &  |\sqrt{2} +1|^2 \approx 6 \nonumber,\\ 
\frac{\Gamma[B^{\pm} \to K^{*\pm} \eta^{\prime}]} {\Gamma[B^{\pm} \to K^{*\pm} \pi^0]}  &\approx &  |\sqrt{2} -1|^2 \approx 
\frac{1}{6}.\
\label{pred}
\eer

Thus, with the new radial admixtures in the wave functions of the
$\eta$ and $\eta^{\prime}$, we can obtain an increase in the $B \to K
\eta^{\prime}$ decay without having the similar increase in the $B \to
K^* \eta^{\prime}$ decay as in the other OZI-violating models
\cite{Atwood}.  In these models additional contributions depend only
on the $\eta^{\prime}$ wave function and are independent of the spin
of the recoiling $K$ or $K^*$;
interference effects do not play the key role of contributing with
opposite signs to $\eta$ and $\eta^{\prime}$ final states as in our
scenario. Thus their predictions contrast sharply with our predicted
suppression for $B \to K^* \eta^{\prime}$ decay as indicated by
Eq.(\ref{pred}) while also predicting enhancement for $B \to K^* \eta$
decay relative to $B \to K^* \pi$, similar to that of $B \to K
\eta^{\prime}$ decay relative to $B \to K \pi$ . It is therefore crucial 
to obtain better
experimental values for the $B \to K^* \eta^{\prime}$, $B \to K^*
\eta$ and $B \to K^* \pi$ decays since no other model gives these
predictions.

We have suggested that the OZI suppressed terms Fig. (\ref{eta}.c) may
play an important role in the decays with $\eta^{\prime}$ in the final
state.
In fact, using Eqs. (\ref{Isgur}, \ref{fsymmetry})
we find for the OZI suppressed contribution,
\ber 
\frac{A^{OZI}_{\eta^{\prime}}K(K^*)} {A^{OZI}_{\eta}K(K^*)} &
\approx & \frac{(1+\frac{1}{\sqrt{2}})} {(1-\frac{1}{\sqrt{2}})} \sim
\frac{A^{OZI}_{\eta^{\prime}}\Lambda} {A^{OZI}_{\eta} \Lambda} \sim
6\ 
\label{OZI} 
\eer 
In particular, the OZI suppressed contribution may be important for
case of $B \to K^* \eta^{\prime}$ decay where there is large
destructive interference between the two leading penguin contributions
Figs.( \ref{eta}.a and \ref{eta}.b).

In Table \ref{T2} we present our full predictions for the decays $
B\rightarrow \eta(\eta^{\prime}) K(K^{*})$ and $\Lambda_b \rightarrow
\Lambda \eta(\eta^{\prime})$ decays.  The expressions for the
amplitudes for $ B$ and $\Lambda_b$ decays can be found in Ref
\cite{Dattaeta} and in Ref \cite{Wafia}, respectively.  These
amplitudes include all terms in the effective Hamiltonian in the
factorization assumption, including the OZI suppressed terms; also
included are chirally enhanced terms which are formally suppressed by
$m_b$ but which come with enhanced coefficients and so are important
for these decays \cite{Etaold}.

The values used for the decay constants are from Eq. (\ref{fsymmetry}) 
 and Eq. (\ref{new})
$f_{\eta}^{u,d}=f_{\eta^{\prime}}^{u,d} \approx  f_{K}/2$ and
$f_{\eta^{\prime}}^{s}=-f_{\eta}^s \approx  f_{K}/\sqrt{2}$
with $f_K=0.160$ GeV.  The form
factors for $B \rightarrow \eta(\eta^{\prime})$ in the ground state
configurations are taken from Ref \cite{BSW} and $SU(3)$ has been used
to relate the $\Lambda_b \rightarrow \Lambda$ form factors to
$\Lambda_b \rightarrow p$ form factors calculated in Ref
\cite{Lambdaff}.
\begin{table}[thb] 
\caption{Branching ratios(BR) for $B \to \eta(\eta')K(K^*)$ decays and
$\Lambda$ decays} 
\begin{center} 
\begin{tabular}{|c|c|c|} 
\hline 
Process & Experimental BR & Theory BR \\ 
\hline 
$B^-\rightarrow K^- \eta{\prime}$ & $(80 ^{+10}_{-9} \pm 7)
\times 10^{-6}$\cite{Richichi:2000kj},
$(70 \pm 8 \pm 5)
\times 10^{-6}$\cite{Babar}
 &$
62\times 10^{-6}$ \\ 
\hline 
$B^-\rightarrow K^- \eta$ & $ < 6.9 \times 10^{-6}$
\cite{Richichi:2000kj} &$ 2.2 \times 10^{-6}$ \\
\hline
$B^-\rightarrow K^{-*} \eta^{\prime}$ & $ < 35\times 10^{-6}$
\cite{Richichi:2000kj} &$ 1.6 \times 10^{-6}$ \\
\hline 
$B^-\rightarrow K^{-*} \eta $ & $(26.4 ^{+9.6}_{-8.2} \pm 3.3)\times
10^{-6}$\cite{Richichi:2000kj} &$ 9 \times 10^{-6}$ \\ 
\hline 
$\Lambda_b \rightarrow \Lambda \eta $ & - &$ 4.6 \times 10^{-6}$ \\ 
\hline 
$\Lambda_b \rightarrow \Lambda \eta^{\prime} $ & - &$ 12 \times 10^{-6}$ \\ 
\hline 
\end{tabular}
\end{center}
\label{T2}
\end{table}
We see from Table \ref{T2}  that our calculations are in reasonable 
agreement with experiment. From this table we find
\ber
R_K & = & 0.035 \nonumber\\
R_{K^*} & = & 5.6\nonumber\\ 
R_{\Lambda} & = & 0.4 \
\label{newratio}
\eer
Note that our prediction for $R_{K^*}$ in Eq. (\ref{newratio}) is
about a factor 6 smaller than in Eq. (\ref{rkstarn}) because the
latter result does not include the contribution of the OZI suppressed
terms nor the chirally enhanced terms. The chirally enhanced terms
contribute with opposite sign to $K$ and $K^*$ final states
\cite{Dattaeta}.  This slightly decreases $B \rightarrow K
\eta^{\prime}$ relative to $B \rightarrow K \eta$ but reduces
substantially the enhancement of $B \rightarrow K^* \eta$ relative to
$B \rightarrow K \eta^{\prime}$. For $R_{\Lambda}$ the effect of the
OZI and the chirally enhanced terms is to decrease the prediction of
$R_{\Lambda}$ relative to the one in Eq. \ref{rlambdan}.

In summary we have considered the effect of radial mixing in the
pseudoscalar systems generated from hyperfine interaction and the
annihilation term.  For the $\eta-\eta^{\prime}$ system we found the
effects of radial mixing can be appreciable and can seriously affect
the branching ratios for the decays, $B \rightarrow
\eta(\eta^{\prime})K(K^*)$, mainly by modifying the $ B \rightarrow
\eta(\eta^{\prime})$ form factors. We found that the effect of radial
mixing, in conjunction with the interference effects between penguin
amplitudes, can give a good description of the existing data on $B
\rightarrow \eta(\eta^{\prime})$ transitions. Moreover we have
predictions for yet unobserved decays that are unique to our model.
We also looked at the decay $\Lambda_b \rightarrow \Lambda
\eta(\eta^{\prime})$ which is dominated by a single amplitude so that
interference effects significant for the $B$ decays are absent here.
Moreover, since no $\Lambda_b \rightarrow \eta(\eta^{\prime})$ form
factors are involved here, the effect of radial mixing is negligible.
We predict only a modest enhancement of $\Lambda_b \rightarrow \Lambda
\eta^{\prime}$ relative to $\Lambda_b \rightarrow \Lambda \eta$ unlike
in the B system; this which is also another unique prediction of our
model.

\centerline{{\bf Acknowledgment}} This work was supported by the
US-Israel Bi-National Science Foundation and by the Natural Sciences
and Engineering Research Council of Canada.

\end{document}